\documentclass[prc,twocolumn,showpacs,floatfix,nofootinbib,preprintnumbers,superscriptaddress,amsmath,amssymb]{revtex4}
\usepackage[latin1]{inputenc}
\usepackage{amsmath}
\usepackage{amsfonts}
\usepackage{amssymb}
\usepackage{CJK}
\usepackage{graphicx}
\usepackage{color}

\newcommand{\qr}{q}
\newcommand{\thalf}{\tfrac{1}{2}}

\begin{document}
\begin{CJK*}{GB}{gbsn}
\title{Approximate restoration of translational and rotational symmetries within the Lipkin method}
\author{Y. Gao (¸ßÔ­)}
\affiliation{Department of Physics, P.O. Box 35 (YFL), FI-40014
University of Jyv\"askyl\"a, Finland}

\author{J. Dobaczewski}
\affiliation{Department of Physics, P.O. Box 35 (YFL), FI-40014
University of Jyv\"askyl\"a, Finland}
\affiliation{Department of Physics, University of York, Heslington, York YO10 5DD, United Kingdom}
\affiliation{Helsinki Institute of Physics, P.O. Box 64, FI-00014 Helsinki, Finland}

\author{P. Toivanen}
\affiliation{Department of Physics, P.O. Box 35 (YFL), FI-40014
University of Jyv\"askyl\"a, Finland}

\begin{abstract}
\begin{description}
\item[Background:]
Nuclear self-consistent mean-field approaches are rooted in the
density functional theory and, through the spontaneous symmetry
breaking mechanism, allow for including important correlations, while
keeping the simplicity of the approach. Because real ground states
should have all symmetries of the nuclear Hamiltonian, these methods
require subsequent symmetry restoration.

\item[Purpose:]
We implement and study Lipkin method of approximate variation after projection
applied to the restoration of the translational or rotational symmetries.

\item[Methods:]
We use Lipkin operators up to quadratic terms in momenta or angular
momenta with self-consistently determined values of the Peierls-Yoccoz
translational masses or moments of inertia, respectively.
Calculations based on Skyrme energy-density functional are performed
for heavy, deformed, and paired nuclei.

\item[Results:]
In deformed nuclei, the Peierls-Yoccoz translational masses along
three different principal-axes directions of the intrinsic system can
be different, which illustrates different widths of the
total-momentum distributions. Numerically, the differences are only
of the order of a few per cent. For the rotational-symmetry
restoration, the Lipkin method compares favorably with the exact
angular-momentum projection, which requires much larger computational
effort.

\item[Conclusions:]
The Lipkin method of translational and rotational symmetry
restoration is a practical low-cost method of determining the
corresponding correlation energies. It allows for a simultaneous
restoration of several symmetries and can be relatively easily
implemented in the standard self-consistent mean-field calculations,
including those required for the parameter adjustments.

\end{description}
\end{abstract}


\maketitle
\end{CJK*}

\section{Introduction}
The self-consistent mean-field method based on nuclear density
functional theory is an approach suitable for large-scale nuclear
structure calculations, see, e.g.,
Refs.~\cite{[Sto03],[Del10],[Erl12],[Erl13],[Agb14],[Gor15]}. An
important advantage of the mean-field method is that a many-body
problem is transformed into a one-body problem, and that the wave
function of the system can be represented by its one-body density
derived from the Kohn-Sham Slater determinant~\cite{[Koh65a]}. The
spontaneous symmetry breaking (SSB)~\cite{[Rin80]} mechanism then
allows for including important correlations, while keeping the
simplicity of the approach. The SSB mechanism introduces
symmetry-broken one-body densities and mean-field states. For
example, by allowing nuclei to deform, quadruple correlations are
included at the expense of broken rotational symmetry. Similarly,
using the Bogoliubov transformation~\cite{[Rin80],[Bla86]}, pairing
correlations are taken into account, whereas the particle number is
no longer conserved.

Although the use of the SSB mechanism is a big success of the
mean-field method, it requires subsequent symmetry
restoration~\cite{[Rin80]}, because real ground states should have
all symmetries of the nuclear Hamiltonian conserved. This is realized
by developing suitable projection methods, that is, by superposing
wave functions with different values of the so-called generator
coordinates~\cite{[Flo75a]}, and thus building wave function with
required conserved symmetries~\cite{[Pei57]}. The projection can be
performed after the variational process of solving the mean-field
equations, which is called projection after
variation~\cite{[Ben03],[Pei57],[Zdu07]}. Another way is to perform
the variation within the subset of projected wave functions, which is
called variation after projection
(VAP)~\cite{[She02],[Ben03],[Sto07]}. Since energies of the projected
states are still functionals of the unprojected densities, the VAP
method also belongs to the class of density functional theories.
However, as compared to projection after variation, computational
cost of VAP is prohibitive, especially when several broken symmetries
have to be simultaneously restored.

In practice, because of the large computational cost of projection
methods, several approximate methods have already been developed.
Examples of these are the center-of-mass and rotational corrections
evaluated after variation~\cite{[Ton00],[Erl11],[Nog64]}. The Lipkin
VAP method~\cite{[Lip60]}, which we employ in this work, is also an
approximate VAP method. Its key idea is to flatten the Hamiltonian so
that the symmetry-conserving states with different quantum numbers
become degenerate. Then, as a superposition, the symmetry-broken
mean-field ground state gives the same energy as the ground state
with the required symmetry conserved. This method has recently been
successfully applied to the translational-symmetry restoration within
the Hartree-Fock (HF) case~\cite{[Dob09g]} and to the particle-number
restoration within the Hartree-Fock-Bogoliubov (HFB)
calculations~\cite{[Wan14]}.

In the present study, we extend the translational-symmetry Lipkin VAP
method of Ref.~\cite{[Dob09g]} to the case of deformed paired nuclei
and fully implement the same method for the rotational-symmetry
restoration. The paper is organized as follows. In Sec.~\ref{secII},
we present a general description of the Lipkin VAP method and we give
some details regarding our implementation of the translational and
rotational symmetry restoration. Results of calculations and
conclusions are presented in Secs.~\ref{secIII} and~\ref{secIV},
respectively.

\section{Theory}\label{secII}
For an operator $\hat{O}$ that commutes with Hamiltonian $\hat{H}$,
the mean-field state $\left| \Phi \right\rangle$
can be written as a superposition of orthonormal eigenstates of $\hat{O}$:
  \begin{equation}\label{eq_dec}
  \left| \Phi \right\rangle = \sum_i a_i \left| O_i \right\rangle.
  \end{equation}
The average energy of state $\left| O_i \right\rangle$,
or the projected energy, is
  \begin{equation}\label{eq_pre}
  E_i=\left\langle O_i \right| \hat{H} \left| O_i \right\rangle.
  \end{equation}
Here, we fix our attention on symmetry operators $\hat{O}$ that have
discrete spectra; however, a generalization to those having continuous
spectra poses no problem.

Using Eqs.~\eqref{eq_dec} and \eqref{eq_pre}, the average of
$\hat{H}$ in the wave function $\left| \Phi \right\rangle$ can be
expressed as
  \begin{equation}\label{eq_avg}
  E=\frac{\left\langle \Phi \right| \hat{H} \left| \Phi \right\rangle}{\left\langle \Phi | \Phi \right\rangle}
  =\frac{\sum_i \left| a_i \right|^2 E_i}{\sum_i \left| a_i \right|^2}.
  \end{equation}
Without any loss of generality, we can assume that for $i=0$ the
eigenvalue of $\hat{O}$ equals zero, $O_0=0$, and that $E_i$ reaches
its minimum at $i=0$. Then one can always define a non-negative
function $f$, such that $f(0)=0$ and
  \begin{equation}\label{eq_eform}
  E_i=E_0 + f\left(O_i\right).
  \end{equation}
Inserting Eq.~\eqref{eq_eform} into Eq.~\eqref{eq_avg}, one obtains
  \begin{eqnarray}
  E   &=& E_0+\frac{\sum_i \left| a_i \right|^2 f\left( O_i \right)}
                   {\sum_i \left| a_i \right|^2}
       =  E_0+\frac{\left\langle \Phi \right| f\left( \hat{O} \right) \left| \Phi \right\rangle}
                   {\left\langle \Phi | \Phi \right\rangle},
  \end{eqnarray}
which gives,
  \begin{eqnarray}\label{eq:06}
  E_0 &=& \frac{\left\langle \Phi \right| \hat{H}-f
                \left( \hat{O} \right) \left| \Phi \right\rangle}
               {\left\langle \Phi | \Phi \right\rangle}.
  \end{eqnarray}
This means that one can obtain the projected energy $E_0$ of the ground state
$\left|i=0\right\rangle$ by applying variational principle to the
flattened Routhian $\hat{H}-f\left(\hat{O}\right)$~\cite{[Lip60]}.
In the following, we refer to $f\left( \hat{O} \right)$ as the Lipkin
operator.

One way to determine the form of $f(\hat{O})$ is to expand it, for
example, as an $N$-rank Taylor series,
\begin{eqnarray}\label{eq_tay}
f\left(\hat{O}\right) & = & \sum_{n=1}^{N}k_{n}\hat{O}^{n},
\end{eqnarray}
and
perform a polynomial fitting. As proposed by Lipkin~\cite{[Lip60]},
the simplest way to do that is by considering kernels of operators
$\hat{O}^{n}$.
Let $\hat{Q}$ be
an operator that commutes with both $\hat{H}$ and $\hat{O}$,
and $\qr$ be a real parameter. Typically, $\hat{Q}$ can be a
generator of the symmetry group related to operator $\hat{O}$.
Inserting the completeness relation $1=\sum_k | O_k \rangle \langle O_k |$
into the relation
  \begin{equation}
  \left\langle \Phi \right| \hat{H} e^{i\qr\hat{Q}} \left| \Phi \right\rangle
  =\sum_{ij}a_i^* a_j \langle  O_i| \hat{H} e^{i \qr\hat{Q}} | O_j \rangle,
  \end{equation}
and using the fact that $\hat{H}$, $\hat{O}$, and $\hat{Q}$
commute with each other, one gets
  \begin{eqnarray}
  \frac{\left\langle \Phi \right| \hat{H} e^{i\qr\hat{Q}} \left| \Phi \right\rangle}
       {\left\langle \Phi \right| e^{i\qr\hat{Q}} \left| \Phi \right\rangle}
  & = &
  \frac{\sum_{i} |a_i|^2  E_i \langle  O_i|e^{i \qr\hat{Q}} | O_i \rangle}
       {\sum_{j} |a_j|^2 \langle  O_j|e^{i \qr\hat{Q}} | O_j \rangle} \nonumber \\\nonumber
  & = &  \frac{\sum_{i} |a_i|^2  \left[E_0+f\left(O_i\right)\right] \langle  O_i|e^{i \qr\hat{Q}} | O_i \rangle}
  {\sum_{j} |a_j|^2 \langle  O_j|e^{i \qr\hat{Q}} | O_j \rangle} \nonumber \\
  & = & E_0+  \frac{\left\langle \Phi \right| f\left(\hat{O}\right) e^{i\qr\hat{Q}} \left| \Phi \right\rangle}
  {\left\langle \Phi \right| e^{i\qr\hat{Q}} \left| \Phi \right\rangle}\nonumber \\
  & = & E_0+  \sum_{n=1}^{N}k_n\frac{\left\langle \Phi \right| \hat{O}^n e^{i\qr\hat{Q}} \left| \Phi \right\rangle}
  {\left\langle \Phi \right| e^{i\qr\hat{Q}} \left| \Phi \right\rangle}.
  \end{eqnarray}
If ${\left\langle \Phi \right| \hat{O}^n e^{i\qr\hat{Q}} \left| \Phi
\right\rangle}/ {\left\langle \Phi \right| e^{i\qr\hat{Q}} \left| \Phi
\right\rangle}$ is not a constant function of $\qr$, then changing the value of $\qr$ one can obtain
different values of kernels of $\hat{O}^n$. Therefore, to determine values of
all parameters $k_n$, that is, to perform the
$N$-rank polynomial fitting, one simply needs to consider $N+1$ different values of $\qr_m$, $m=0,\ldots,N$,
with $\qr_0=0$,
and solve the problem of $N+1$ linear equations,
  \begin{eqnarray}
  \label{eq:10}
  \frac{\left\langle \Phi \right| \hat{H} e^{i\qr_m\hat{Q}} \left| \Phi \right\rangle}
       {\left\langle \Phi \right| e^{i\qr_m\hat{Q}} \left| \Phi \right\rangle}
  & = & \sum_{n=0}^{N}k_n\frac{\left\langle \Phi \right| \hat{O}^n e^{i\qr_m\hat{Q}} \left| \Phi \right\rangle}
  {\left\langle \Phi \right| e^{i\qr_m\hat{Q}} \left| \Phi \right\rangle},
  \end{eqnarray}
where $k_0\equiv E_0$.

When Taylor expansion (\ref{eq_tay}) is reduced to the first
power only ($N=1$), the 2$\times$2 equation~(\ref{eq:10}) can be
easily solved~\cite{[Dob09g]}, which gives,
  \begin{eqnarray}
  \label{eq:10a} \!\!\!\!\!\!\!\!
  k &=&
  \frac{\left\langle \Phi \right| \hat{H} e^{i\qr\hat{Q}} \left| \Phi \right\rangle
        \left\langle \Phi                                      | \Phi \right\rangle
       -\left\langle \Phi \right| \hat{H}                 \left| \Phi \right\rangle
        \left\langle \Phi \right|         e^{i\qr\hat{Q}} \left| \Phi \right\rangle}
       {\left\langle \Phi \right| \hat{O} e^{i\qr\hat{Q}} \left| \Phi \right\rangle
        \left\langle \Phi                                      | \Phi \right\rangle
       -\left\langle \Phi \right| \hat{O}                 \left| \Phi \right\rangle
        \left\langle \Phi \right|         e^{i\qr\hat{Q}} \left| \Phi \right\rangle},
  \end{eqnarray}
where $k\equiv{}k_1$ and $\qr\equiv{}\qr_1$.
However, even when a few terms are
kept in the Taylor expansion~\cite{[Wan14]}, the solution poses no
problem either. Similarly, when the Lipkin operator contains not one
but several operators $\hat{O}$, the combined linear equations will
still have quite low dimensions. In practical cases,
equations pertaining to different operators may turn out to be fairly
well decoupled, and can be solved independently. This is, in fact,
the case in our practical applications to translational-symmetry
restoration, where shifts in three Cartesian directions are well
decoupled from one another, and can be treated separately.

When $\hat{O}$ is chosen to be $\hat{Q}^2$, it is possible to use the
Gaussian overlap approximation (GOA)~\cite{[Rin80]} to obtain estimates of
kernels $\left\langle \Phi \right| \hat{Q}^{2n}
e^{i\qr\hat{Q}}\left| \Phi \right\rangle/ \left\langle \Phi
\right|e^{i\qr\hat{Q}}\left| \Phi \right\rangle$. Within the GOA, we have
  \begin{equation}\label{eq_goa}
  \left\langle \Phi \right| e^{i \qr\hat{Q} }\left| \Phi \right\rangle
  \approx e^{-\frac{1}{2}a \qr^2},\ a \equiv \left\langle \Phi \right| \hat{Q} ^2\left| \Phi \right\rangle.
  \end{equation}
From this, one can find that
  \begin{eqnarray}
  \frac{\left\langle \Phi \right| \hat{Q}^{2n}e^{i \qr\hat{Q} }\left| \Phi \right\rangle}
           {\left\langle \Phi \right| e^{i \qr\hat{Q} }\left| \Phi \right\rangle}
  & = &
  (-)^n
   \frac{\frac{d^{2n}}{d \qr^{2n}}
        {\left\langle \Phi \right| e^{i \qr\hat{Q} } \left| \Phi \right\rangle}}
        {\left\langle \Phi \right| e^{i \qr\hat{Q} } \left| \Phi \right\rangle}
 \nonumber \\
   \approx
  (-)^n e^{\frac{1}{2}a \qr^2}
        \frac{d^{2n}}
             {d \qr^{2n}} e^{-\frac{1}{2}a \qr^2}
  & = &
  \left(-\tfrac{a}{2}\right)^nH_{2n}\left(\sqrt{\tfrac{a}{2}} \qr\right),
  \end{eqnarray}
where $H_{2n}$ is the Hermite polynomial of order $2n$. In
addition, from expression \eqref{eq_goa}, $a$ can be calculated
from the following formula,
  \begin{equation}
  a\approx-2\lim_{\qr\rightarrow0}\frac{\ln(\left\langle \Phi \right| e^{i \qr\hat{Q} }\left| \Phi \right\rangle)}{\qr^2}.
  \end{equation}

In this work, for the translational-symmetry restoration,
$f\left(\hat{O}\right)$, $\hat{Q}$, and $\qr$ are chosen as:
  \begin{equation}
  \label{eg:14}
  f=\sum_{i=x,y,z} k_i \hat{P}_i^2,\;\;\hat{Q}=\hat{P}_{i=x,y,z},\;\;
     \qr=\delta{}x,\delta{}y,\delta{}z,
  \end{equation}
where $\hat{P}_i$ are components of the total momentum operator in
three Cartesian directions. In this way, for $i=x,y,z$, operator
$e^{i \qr\hat{Q}}$ shifts the nucleus by the distance of
$\delta{}x,\delta{}y,\delta{}z$ in the direction of the $x,y,z$ axis,
respectively. Three different operators $\hat{Q}$ are used to make
sure that kernels of squares of the total momentum in each direction,
${\left\langle \Phi \right| \hat{P}^2_i e^{i\qr\hat{Q}} \left| \Phi
\right\rangle}/ {\left\langle \Phi \right| e^{i\qr\hat{Q}} \left|
\Phi \right\rangle}$, do really change with shifts $\qr$.

The form of the Lipkin operator $f(\hat{O})$ adopted for the
translational-symmetry restoration is motivated by the following
arguments. First, we expect that the energy expectation value has a
similar form as the energy of the center-of-mass motion, which is
$\mathbf{\hat{P}}^2/2M$~\cite{[Lip60]}. However, since we consider
deformed nuclei, dispersion of the total momentum can be anisotropic,
and thus the three Lipkin parameters $k_i$ can be different. Second,
terms linear in momentum should disappear, because we study
ground states of even-even nuclei and thus the time-reversal symmetry
is not broken. Third, higher order terms are expected to be very
small~\cite{[Dob09g]}, and anyhow their implementation would have been
very complicated.

For the rotational-symmetry restoration, in this paper we
concentrate on axially deformed even-even nuclei.
Therefore,
$f\left(\hat{O}\right)$, $\hat{Q}$, and $\qr$ are now chosen as:
  \begin{equation}\label{eq_kofj}
  f=k \left({\hat{J}}_{x}^2+{\hat{J}}_{y}^2\right)
    ,\;\; \hat{Q}=\hat{J}_{y},\;\; \qr=\beta,
  \end{equation}
where the axial-symmetry axis is aligned with the Cartesian $z$
direction, $\hat{J}_{x}$ and $\hat{J}_{y}$ are the $x$ and $y$
components of the total angular momentum, respectively, and $\beta$
is the Euler rotation angle about the $y$-axis.\footnote{Depending on
the context, the same symbol $\beta$ is traditionally used both for
the second Euler angle and Bohr deformation, which must not be
confused.} Here, operator $e^{i \qr\hat{Q}}$ rotates the nucleus by
angle $\beta$ around the $y$ axis. As in the translational case, because
of the conserved time-reversal symmetry, linear terms are dropped.
However, the fact that we neglect here higher-order terms restricts
the present application to the case of collective rotation with
energy increasing as square of the total angular momentum. By the
same token, since the energy of an axial nucleus does not increase
with increasing rotation about the symmetry axis, in the Lipkin
operator we dropped the term ${\hat{J}}_{z}^2$. Anyhow, in the ground
state of an even-even axial nucleus, the average value of this term
vanishes, $\left\langle\hat{J}_z^2\right\rangle = 0$.

\section{Results and Discussion}\label{secIII}

Implementations of the approximate restoration of translational and
rotational symmetries within the Lipkin method were here carried out
using the code {\sc hfodd} (v2.73m)~\cite{[Sch12],[Sch14]}, which solves the HFB
equations on a 3D Cartesian harmonic oscillator basis. All
calculations were performed in the space of 16 major spherical
harmonic-oscillator shells and with the Skyrme SLy4
parametrization~\cite{[Cha98a]}. For the HFB calculations, a volume
zero-range pairing interaction with a cutoff window of
$E_{\mathrm{cut}}=60$\,MeV was used. For the translational case, we
only performed calculations with pairing, and the pairing strengths
were adjusted to reproduce experimental odd-even mass staggering in
the entire rare-earth region, which gave $V_n=-159$\,MeV\,fm$^3$ for
neutrons and $V_p=-152$\,MeV\,fm$^3$ for protons. For the rotational
case, we performed calculations both with and without pairing
correlations included. Here, the pairing strengths were adjusted to
reproduce experimental odd-even mass staggering in nuclei around
$^{168}$Er, which gave $V_n=-202.17$ and $V_p=-221.70$\,MeV\,fm$^3$.
To improve the convergence when solving the HFB equations, the
two-basis method~\cite{[Gal94],[Sch12]} was used.

\subsection{Restoration of the translational symmetry}\label{secIIIa}

As discussed in Ref.~\cite{[Dob09g]}, coefficients $k_i$ that define
the Lipkin operator (\ref{eg:14}) correspond to the so-called
Peierls-Yoccoz masses~\cite{[Pei57]} and characterize the momentum
contents of the symmetry-broken stationary ground-state wave function.
Conversely, the exact inertial mass $mA$ characterizes the reaction of
the system to the boost (an increase in momentum). As it turns out,
both masses have similar but not identical values~\cite{[Dob09g]}. In
the present work we represent the calculated values of $k_i$ through
the ratios of the corresponding Peierls-Yoccoz and exact masses, that
is,
  \begin{equation}\label{eq:15}
R_i = \frac{M_{\text{PY},i}}{mA} = \frac{1}{2k_imA}\quad\mbox{for}\quad i=x,y,z.
  \end{equation}
Needless to say that in deformed nuclei, the momentum contents in the
three Cartesian directions of the intrinsic system can be different,
and thus the three Peierls-Yoccoz masses $M_{\text{PY},i}$ can be
different too.

\begin{figure}[htbp]
  \centering
  \includegraphics[width=0.45\textwidth]{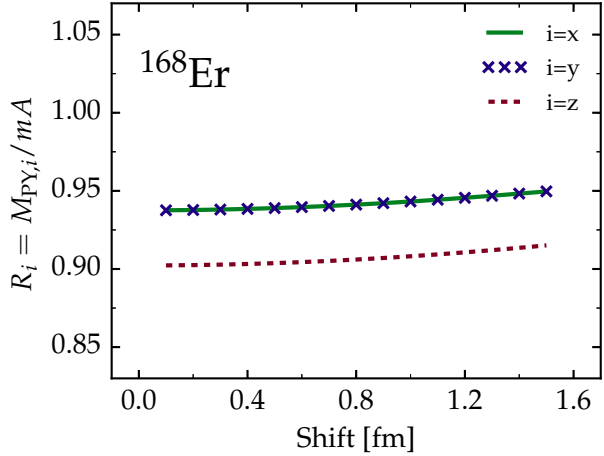}
  \caption{Ratios (\protect\ref{eq:15}) between the Peierls-Yoccoz
           masses in the $x$, $y$, and $z$ directions and the
           exact mass as functions of shifts in the
           corresponding directions, calculated for $\mathrm{^{168}Er}$.
}
  \label{fig_mri}
  \end{figure}

The effectiveness of the chosen Lipkin operator $f(\hat{O})$
(\ref{eg:14}), used in the translational-symmetry restoration, can be
tested by checking an eventual dependence of the Lipkin parameters
$k_{x,y,z}$ on shifts $\delta{}x,\delta{}y,\delta{}z$. In
Fig.~\ref{fig_mri} we show this dependence for $\mathrm{^{168}Er}$.
One can see that when the shifts change from 0.05 to 1.5\,fm, the
Lipkin mass does not change much. This indicates that the quadratic
shape is already a good approximation of the total momentum contents
of the HFB wave function. In the following calculations, we fix
the shifts at 0.5\,fm.

  \begin{figure}[htbp]
        \centering
        \includegraphics[width=0.45\textwidth]{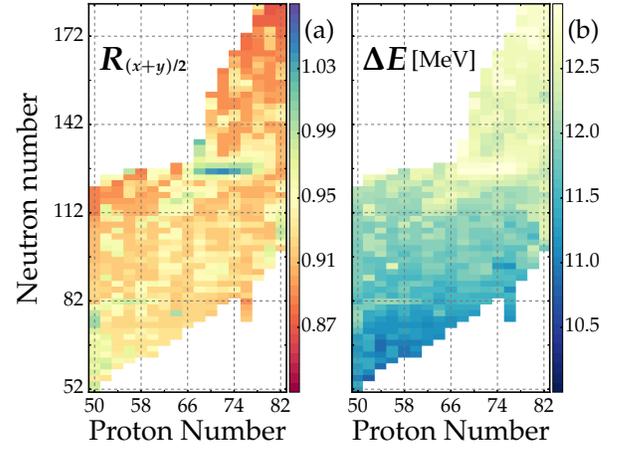}
        \caption{
         (a) Average ratios of the
         Peierls-Yoccoz and exact masses in the $x$ and $y$ directions
         (\protect\ref{eq:16}).
         (b) Differences $\Delta{E}$ between the Lipkin VAP energies
         and those obtained within the standard HFB calculations.}
        \label{fig_m1}
  \end{figure}

  \begin{figure}[htbp]
        \centering
        \includegraphics[width=0.45\textwidth]{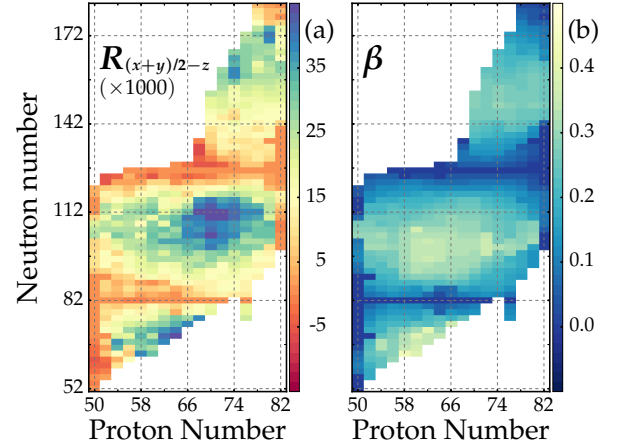}
        \caption{(a) $R_{\left(x+y\right)/2}-R_{z}$: The difference between $R_{\left(x+y\right)/2}$ and
                the mass ratio in the $z$-direction.
                (b) Beta deformations.}
        \label{fig_m2}
  \end{figure}

    \begin{figure}[htbp]
        \centering
        \includegraphics[width=0.45\textwidth]{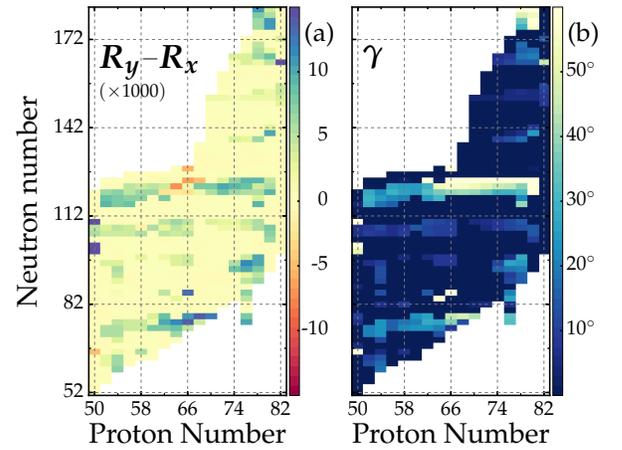}
        \caption{(a) $R_{y}-R_{x}$:
                The difference between the mass ratios in the $y$- and $x$-directions.
                (b) Gamma deformations.}
        \label{fig_m3}
    \end{figure}

In Fig.~\ref{fig_m1}(a), we plotted the average values of $R_x$ and $R_y$,
  \begin{equation}\label{eq:16}
  R_{(x+y)/2}  \equiv \thalf\left(R_x+R_y\right),
  \end{equation}
which characterize the overall values of the Peierls-Yoccoz masses
in the directions perpendicular to the symmetry axis. As we can see,
the Peierls-Yoccoz masses vary fairly smoothly with particle numbers
and are almost unaffected by deformations of nuclei. Similarly, the
Lipkin VAP energies (\ref{eq:06}), Fig.~\ref{fig_m1}(b), smoothly
overestimate the standard HFB energies, where the one-body center-of-mass
corrections have been used~\cite{[Ben00d],[Dob09g]}.

In Figs.~\ref{fig_m2}(a) and~\ref{fig_m3}(a), to illustrate the
anisotropy of the Peierls-Yoccoz masses, we plotted
differences of ratios (\ref{eq:15}),
  \begin{equation}\label{eq:17}
  R_{(x+y)/2-z}  \equiv \thalf\left(R_x+R_y\right)-R_z,
  \end{equation}
and $R_y-R_x$, respectively.
First we note that the overall degree of the anisotropy is fairly
small, with the above differences not exceeding a few per cent.
Differences (\ref{eq:17}) characterize the anisotropy between the
directions perpendicular and parallel to the symmetry axis,
and very nicely correlate with values of deformations $\beta$,
plotted in Fig.~\ref{fig_m2}(b). Similarly, differences $R_y-R_x$
correlate with the nonaxiality of shapes, as illustrated by values
of deformations $\gamma$, plotted in Fig.~\ref{fig_m3}(b).

\subsection{Restoration of the rotational symmetry}\label{secIIIb}

\begin{figure}[htbp]
  \centering
  \includegraphics[width=0.45\textwidth]{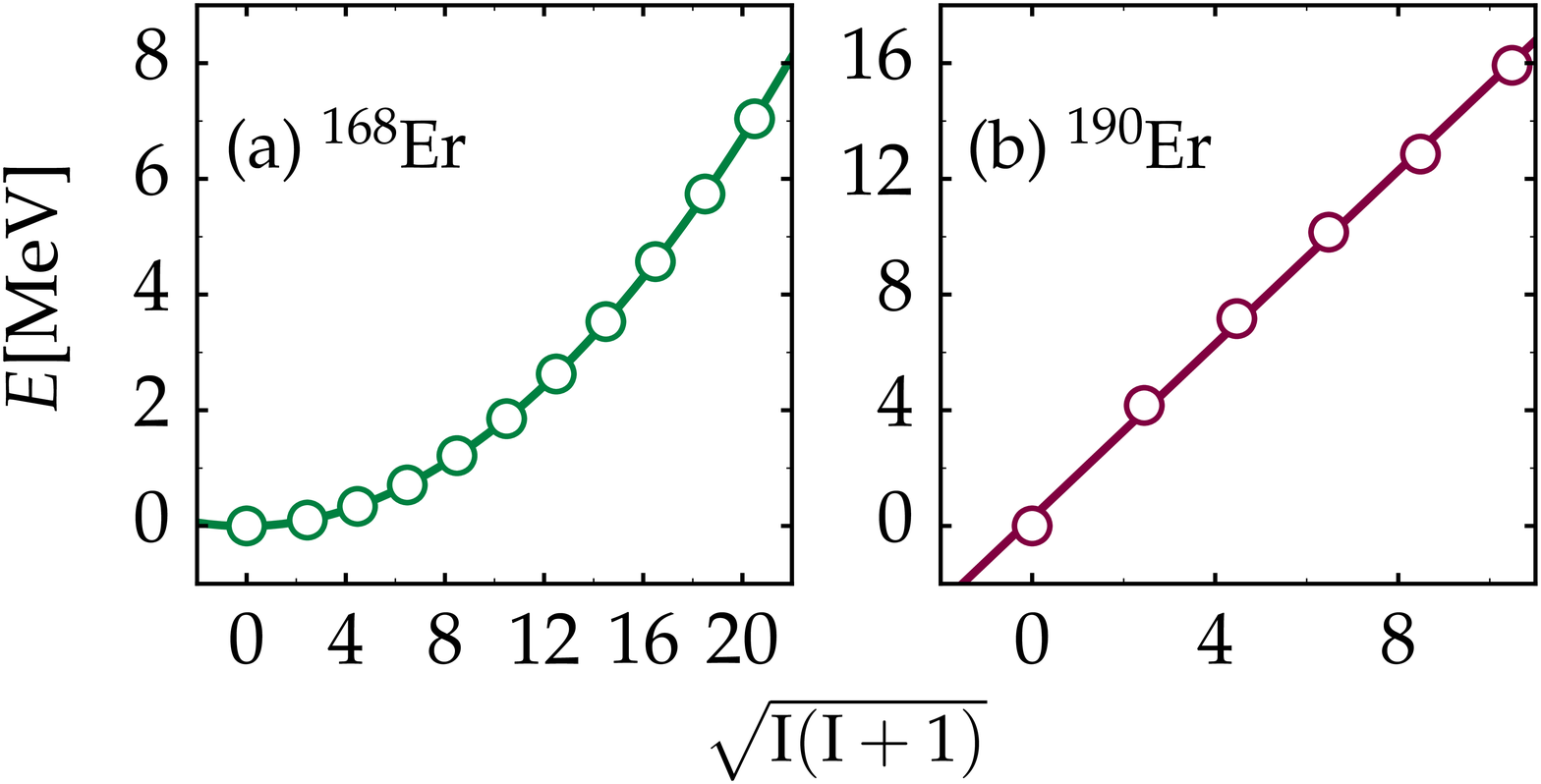}
  \caption{Spectra of (a) $\mathrm{^{168}Er}$ and (b) $\mathrm{^{190}Er}$. The angular momentum projection
           was performed after the convergence of the standard HFB calculation.}
  \label{fig_epro}
  \end{figure}

We begin by showing, in Fig. \ref{fig_epro}, spectra of
well-deformed, $\mathrm{^{168}Er}$, and weakly-deformed,
$\mathrm{^{190}Er}$, nucleus, calculated using the angular-momentum
projection (AMP) of the standard HFB solutions. The results indicate
a rotational mode of the former and a vibrational mode of the
latter~\cite{[Rin80]}. This means that in weakly-deformed nuclei, the
parabolic Lipkin operator $f(\hat{O})$ (\ref{eq_kofj}) may not be
appropriate to flatten the total energy. Guided by these results,
below we consider only well-deformed erbium isotopes with
$N=86$--118, with some weakly-deformed nuclei still included for
possible further comparisons.

\begin{figure}[htbp]
  \centering
  \includegraphics[width=0.45\textwidth]{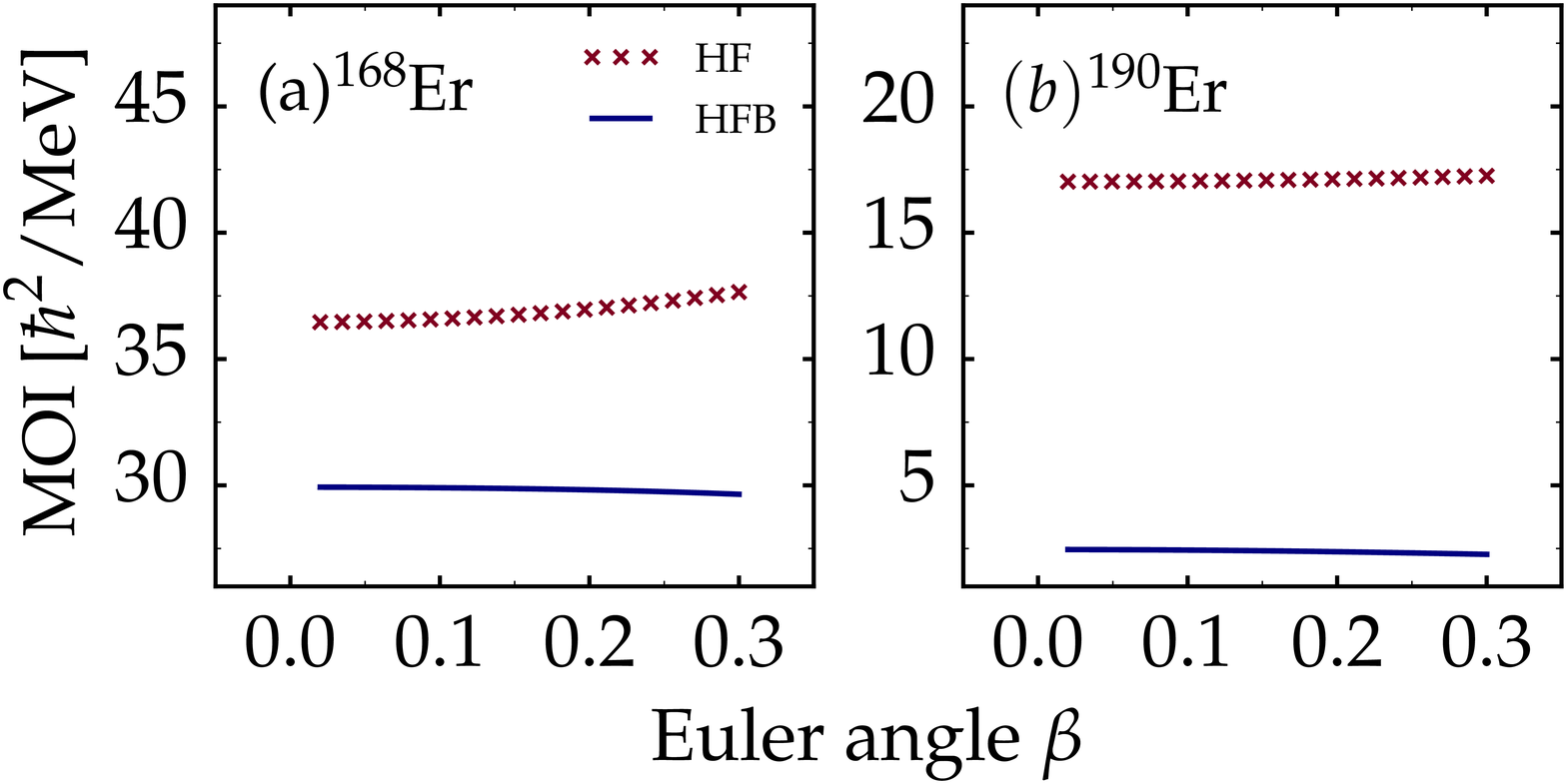}
  \caption{The Lipkin MoI (\protect\ref{eq:18}) determined for (a) $\mathrm{^{168}Er}$ and (b) $\mathrm{^{190}Er}$,
   plotted as functions of the Euler rotation angle $\beta$.}
  \label{fig_ibeta}
  \end{figure}

All results below are shown for the
Lipkin parameters $k$ re-represented in terms of the Lipkin Moments of Inertia (MoI)
${\cal J}$ as
  \begin{equation}\label{eq:18}
  {\cal J}  \equiv \frac{1}{2k}.
  \end{equation}
Similarly as in the case the translational symmetry, we begin by
checking the dependence of results on the Euler rotation angle
$\beta$. Since in well-deformed and weakly-deformed nuclei, the
dependence of total energy on total angular momentum is different, in
Fig.~\ref{fig_ibeta} we show the MoI determined in
$\mathrm{^{168}Er}$ and $\mathrm{^{190}Er}$ as functions of the Euler
rotation angle $\beta$. For both nuclei, we show results obtained
with and without pairing correlations included. We found that in all
cases, the Lipkin MoI very weakly depend on $\beta$. Therefore, results
presented below were obtained using $\beta=0.1$.

\begin{figure}[htbp]
  \centering
  \includegraphics[width=0.45\textwidth]{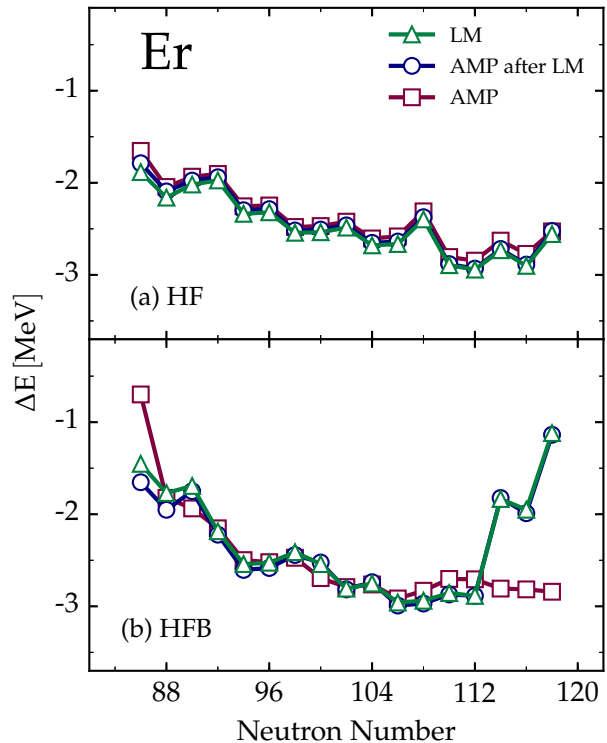}
  \caption{Energy corrections obtained by the Lipkin VAP method (LM),
   AMP, and AMP after LM
  for erbium isotopes (a) without or (b) with pairing correlations.}
  \label{fig_ecorr}
  \end{figure}

In Fig.~\ref{fig_ecorr}, we show energy corrections obtained by the
Lipkin VAP, AMP after the convergence of the standard mean-field
calculations, and AMP after the convergence of the Lipkin VAP
calculations. In the case of calculations without pairing,
Fig.~\ref{fig_ecorr}(a), the results obtained by all three methods
are quite similar. With pairing, Fig.~\ref{fig_ecorr}(b), the same is
true for results obtained by the Lipkin VAP and AMP after Lipkin VAP.
A general agreement between the results of Lipkin VAP and AMP after
Lipkin VAP supports the validity of the Lipkin VAP method as an
approximation of the exact VAP method.

\begin{figure}[htbp]
  \centering
  \includegraphics[width=0.45\textwidth]{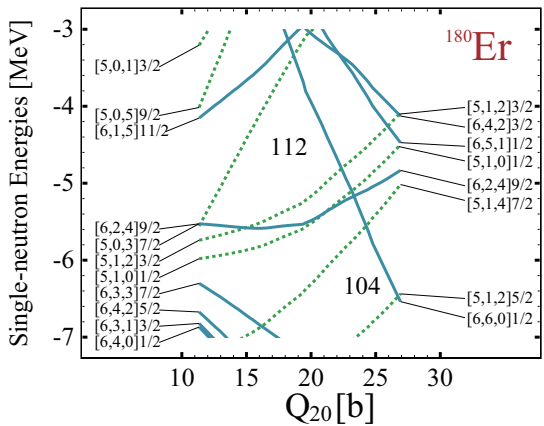}
  \caption{Neutron Nilsson diagram for $\mathrm{^{180}Er}$.
   Single-particle levels were determined as eigenenergies of the mean fields
   obtained using the deformation-constrained HFB calculations.}
  \label{fig_nil}
  \end{figure}

\begin{figure}[htbp]
  \centering
  \includegraphics[width=0.45\textwidth]{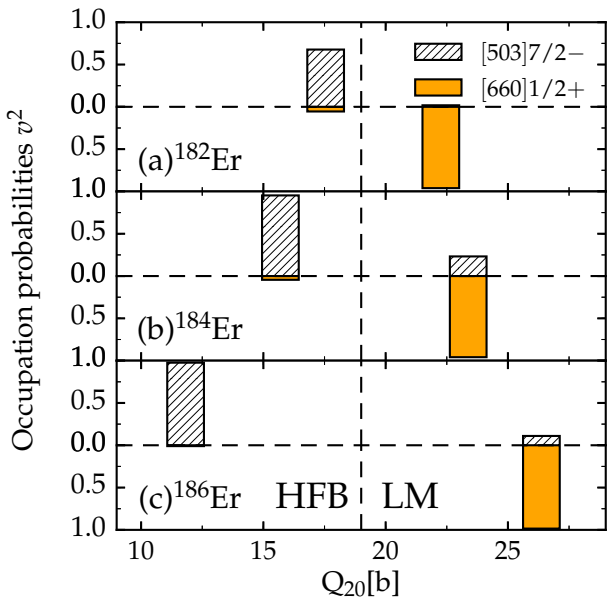}
  \caption{Occupation probabilities $v^2$ of neutron orbitals
     $\nu\left[660\right]1/2+$
     and $\nu\left[503\right]7/2-$
  in (a) $\mathrm{^{182}Er}$,
     (b) $\mathrm{^{184}Er}$, and
     (c) $\mathrm{^{186}Er}$, determined within the
     HFB (left panel) and Lipkin (right panel) methods.}
  \label{fig_occ}
  \end{figure}

In $\mathrm{^{154,182-186}Er}$, owing to changes in deformation,
marked differences appear between the standard and Lipkin VAP calculations
with pairing. For example, as seen in the neutron Nilsson diagram for
$\mathrm{^{180}Er}$, Fig.~\ref{fig_nil}, the intruder orbit
$\nu\left[660\right]\frac{1}{2}+$ crosses the extruder orbit
$\nu\left[503\right]\frac{7}{2}-$. Thus in $\mathrm{^{182-186}Er}$,
the competition between the two configurations leads to different
shapes obtained in the standard HFB and Lipkin methods. This is
substantiated in Fig.~\ref{fig_occ}, where we plotted occupation
probabilities $v^2$ of these two orbits in $\mathrm{^{182-186}Er}$. By
occupying the intruder orbit, the nuclei are driven to larger
deformations.

\begin{figure}[htbp]
        \centering
        \includegraphics[width=0.45\textwidth]{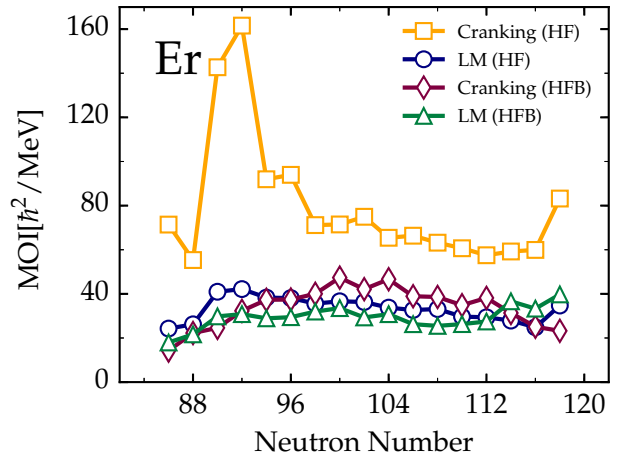}
        \caption{MoI obtained using the Lipkin VAP method (LM) and cranking method
         with pairing correlations included (HFB) or not included (HF).
        }
        \label{fig_moi}
  \end{figure}

In Fig.~\ref{fig_moi}, we compare the MoI calculated within the Lipkin VAP with
those determined using the cranking method.
The cranking calculations were performed at the frequency of $\omega=0.05$\,MeV
and the first MoI were extracted as ratios of average angular momentum and frequency.
Without pairing, the cranking MoI are significantly
larger than those obtained using the Lipkin VAP. Once again, this is because
they are two different quantities, corresponding to the
Thouless-Valatin and Peierls-Yoccoz MoI, respectively~\cite{[Rin80]}.
The former illustrate how the system reacts to rotation, whereas the
latter characterize average energies of components with good total
angular momentum within a non-rotating broken-symmetry ground state.
As it turns out, when the pairing is included, differences between
the cranking and Lipkin VAP MoI become much smaller.

\section{Conclusions}\label{secIV}

In the present study, we applied the Lipkin method of the
translational and rotational symmetry restoration so as to
approximate the exact variation-after-projection method.
We implemented the Lipkin method by constructing corrective
Lipkin operators up to quadratic terms in linear and angular momenta,
and we self-consistently determined the corresponding Peierls-Yoccoz
translational masses and moments of inertia, respectively.

For the translational symmetry restoration, we performed calculations
for even-even isotopes of elements with $50\leq{Z}\leq82$. We found
that the Peierls-Yoccoz masses in different principal-axes directions
of the intrinsic system differ up to a few per cent, which
illustrates differences in linear-momentum distributions in deformed
nuclei. For the rotational symmetry restoration, we performed calculations
for even-even erbium isotopes. Here we found that the correlation
energies obtained within the Lipkin method nicely reproduce results
of the exact angular-momentum projection.

\begin{acknowledgments}
This work was supported in part
by the Academy of Finland and University of Jyv\"askyl\"a within the FIDIPRO program,
by the Polish National Science Center under Contract No.\ 2012/07/B/ST2/03907, and
by the ERANET-NuPNET grant SARFEN of the Polish National Centre for Research and Development (NCBiR).
We acknowledge the CSC-IT Center for Science Ltd., Finland, for the allocation of
computational resources.
\end{acknowledgments}

\bibliographystyle{unsrt}

\end{document}